\documentclass[prb,aps,superscriptaddress,twocolumn,dvips]{revtex4}
\usepackage{feynmp}
\usepackage{float}
\usepackage{amsmath}
\usepackage{amsfonts}
\usepackage{amssymb}
\usepackage{epsfig}
\usepackage{graphicx}
\usepackage{subfigure}
\usepackage{bbm}
\usepackage{color}
\usepackage{longtable}
\usepackage{booktabs}
\usepackage{multirow}
\usepackage{graphicx}
\usepackage{epsfig}
\usepackage{bm}
\usepackage{txfonts}

\newcommand{\Rmnum}[1]{\expandafter\@slowromancap\romannumeral #1@}
\allowdisplaybreaks[3]
\begin{document}

\title{Fermion dispersion renormalization in a two-dimensional semi-Dirac semimetal}

\author{Hao-Fu Zhu}
\affiliation{CAS Key Laboratory for Research in Galaxies and
Cosmology, Department of Astronomy, University of Science and
Technology of China, Hefei, Anhui 230026, China}\affiliation{School
of Astronomy and Space Science, University of Science and Technology
of China, Hefei, Anhui 230026, China}
\author{Xiao-Yin Pan}
\affiliation{Department of Physics, Ningbo University, Ningbo,
Zhejiang 315211, China}
\author{Guo-Zhu Liu}
\altaffiliation{Corresponding author: gzliu@ustc.edu.cn}
\affiliation{Department of Modern Physics, University of Science and
Technology of China, Hefei, Anhui 230026, China}

\begin{abstract}
We present a non-perturbative study of the quantum many-body effects
caused by the long-range Coulomb interaction in a two-dimensional
semi-Dirac semimetal. This kind of semimetal may be realized in
deformed graphene and a class of other realistic materials. In the
non-interacting limit, the dispersion of semi-Dirac fermion is
linear in one direction and quadratic in the other direction. When
the impact of Coulomb interaction is taken into account, such a
dispersion can be significantly modified. To reveal the correlation
effects, we first obtain the exact self-consistent Dyson-Schwinger
equation of the full fermion propagator and then extract the
momentum dependence of the renormalized fermion dispersion from the
numerical solutions. Our results show that the fermion dispersion
becomes linear in two directions. These results are compared to
previous theoretical works on semi-Dirac semimetals.
\end{abstract}

\maketitle

%%%%%%%%%%%%%%%%%%%%%%%%%%%%%Main Body%%%%%%%%%%%%%%%%%%%%%%%%%

\section{Introduction \label{Sec:Introdunction}}

Over ten types of semimetal materials have been discovered
\cite{CastroNeto, Kotov12, Vafek14, Hirata21} in the past two
decades. These materials exhibit a variety of unusual phenomena that
cannot be observed in ordinary metals and as such have attracted
intense theoretical and experimental investigations. Although most
research works are concentrated on the single-particle properties of
semimetals, such as nontrivial topology and chiral anomaly, under
proper conditions the inter-particle interactions may play a vital
role \cite{CastroNeto, Kotov12, Vafek14, Hirata21}, leading to
strong renormalization of various physical quantities and also some
ordering instabilities.

The Coulomb interaction is usually unimportant in ordinary metals
that have a finite Fermi surface. The reason of this fact is that
the originally long-range Coulomb interaction becomes short-ranged
due to the static screening induced by the finite density of states
(DOS) of electrons on the Fermi surface. Renormalization group (RG)
analysis indicates that weak short-ranged repulsion is an irrelevant
perturbation at low energies \cite{Shankar}. Direct perturbative
calculations \cite{Coleman} show that short-ranged Coulomb
interaction only leads to weak renormalization of physical
quantities, which ensures the validity of Landau's Fermi liquid
theory in ordinary metals. The role of Coulomb interaction could be
rather different in semimetals that host band-touching points. Let
us take two-dimensional (2D) Dirac semimetal (DSM) as an example. In
the non-interacting limit, undoped DSM manifests a perfect Dirac
cone near the band-touching neutral point \cite{CastroNeto,
Kotov12}. Dirac fermions emerges at low energies with a linear
dispersion and a constant velocity $v_{F}$. Since the DOS of Dirac
fermions vanishes at the band-touching point, the Coulomb
interaction remains long-ranged and can lead to considerable quantum
many-body effects \cite{CastroNeto, Kotov12}. For instance, the
fermion velocity acquires a logarithmic dependence on momentum owing
to the Coulomb interaction \cite{CastroNeto, Kotov12, Gonzalez94,
Elias11, Lanzara11, Chae12, Pan21}. As a consequence, the original
Dirac cone is apparently reshaped near the band-touching point
\cite{Elias11}.

\begin{figure}[H]
\centering
\includegraphics[width=3.26in]{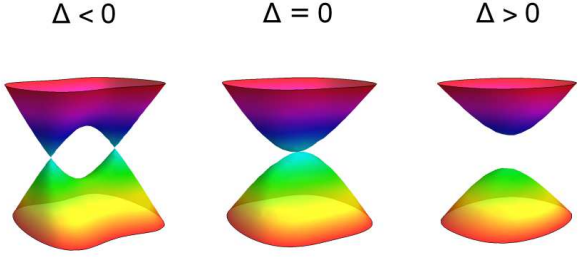}
\caption{The quantum phase transition between a DSM and a BI is
tuned by changing the sign of the gap $\Delta$. For $\Delta<0$, the
system is a DSM with two band-touching Dirac points. As
$\Delta\rightarrow 0$, two isolated Dirac points merge into one
single point at which 2D semi-Dirac fermions emerge. If $\Delta>0$,
the system becomes a normal BI.} \label{semi-insu}
\end{figure}

One can tune some parameters of a 2D DSM to drive two separate Dirac
points to merge into one single point \cite{Montambaux09A,
Montambaux09B, Bellec13, Lim12}. Such a manipulation generates a new
type of 2D semimetal. This new semimetal is called semi-DSM in the
literature since the fermion dispersion is linear along one (say,
$x$-) axis and quadratic along the other (correspondingly, $y$-)
axis. The kinetic energy of 2D semi-Dirac fermions is expressed as
\begin{eqnarray}
E = \pm\sqrt{\upsilon^{2}p_{x}^{2}+B^{2}p_{y}^{4}},
\end{eqnarray}
which $\upsilon$ is the effective velocity in the $x$-direction and
$B$ stands for the inverse of the effective mass in the
$y$-direction. These fermions are the low-energy elementary
excitations at the quantum critical point (QCP) of the phase
transition from a 2D DSM to a band insulator (BI), as illustrated in
Fig.~\ref{semi-insu}. Such kind of semi-Dirac fermions could be
realized in a variety of materials, including deformed graphene
\cite{Dietl08, Montambaux09A, Montambaux09B}, pressured organic
compound $\alpha$-(BEDT-TTF)$_{2}$I$_{3}$ \cite{Goerbig08,
Montambaux09B, Kobayashi07, Kobayashi11}, certain TiO$_{2}$/VO$_{2}$
nano-structures \cite{Pardo09, Pardo10, Banerjee09}, properly doped
few-layer black phosphorus \cite{Kim15}, and some artificial optical
lattices \cite{Wunsch08, Tarruell12}. In the past ten years,
considerable research efforts have been devoted to studying the
interaction-induced correlation effects \cite{Isobe16, Cho16,
Kotov21}, the hydrodynamic transport properties \cite{Schmalian18},
and also several possible ordering instabilities \cite{Wang17,
Uchoa17, Uchoa19, Roy18} in 2D semi-DSMs.

Similar to 2D DSM, the DOS of the semi-Dirac fermions vanishes at
the Fermi level. Thus the Coulomb interaction is also long-ranged.
It is important to examine whether the poorly screened Coulomb
interaction produces nontrivial correlation effects. This issue has
been previously addressed by several groups of theorists. Isobe
\emph{et al.} \cite{Isobe16} carried out RG calculations based on
the $1/N$ expansion, where $N$ is the fermion flavor, and claimed to
reveal a crossover from non-Fermi liquid behavior to marginal Fermi
liquid behavior as the energy scale is lowered. Cho and Moon
\cite{Cho16} made a RG analysis of the same model system by
employing two different perturbative expansion schemes and predicted
the existence of a novel quantum criticality characterized by the
anisotropic renormalization of the Coulomb interaction. Recently,
Kotov \emph{et al.} \cite{Kotov21} re-visited this problem. They
employed the weak-coupling expansion method to perform leading-order
RG calculations, choosing the fine structure constant $\alpha$ as a
small parameter, and found a weak-coupling fixed point that appears
to be different from the unusual fixed points obtained by means of
$1/N$ expansion \cite{Isobe16, Cho16}. In particular, Kotov \emph{et
al.} \cite{Kotov21} showed that the fermion dispersion becomes
linear in both the $x$- and $y$-directions after incorporating the
renormalization caused by the Coulomb interaction.

\begin{figure}[H]
\centering \subfigure[]{\label{Fig.sub.1}
\includegraphics[width=1.2in]{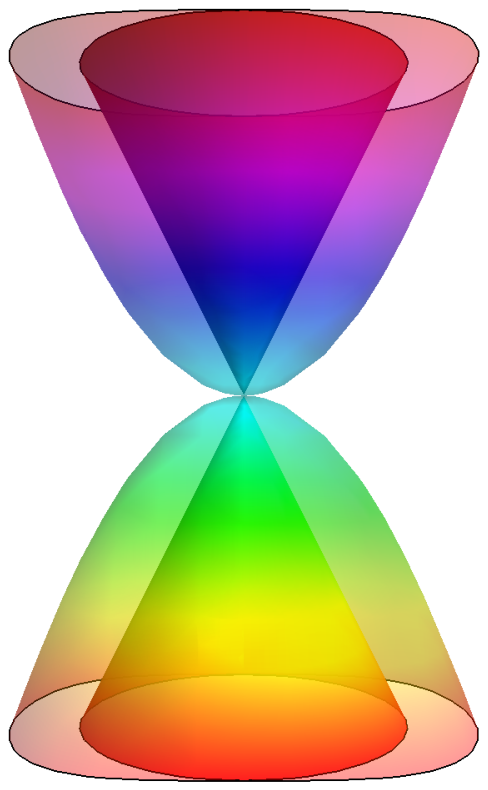}}
\subfigure[]{\label{Fig.sub.2}
\includegraphics[width=1.26in]{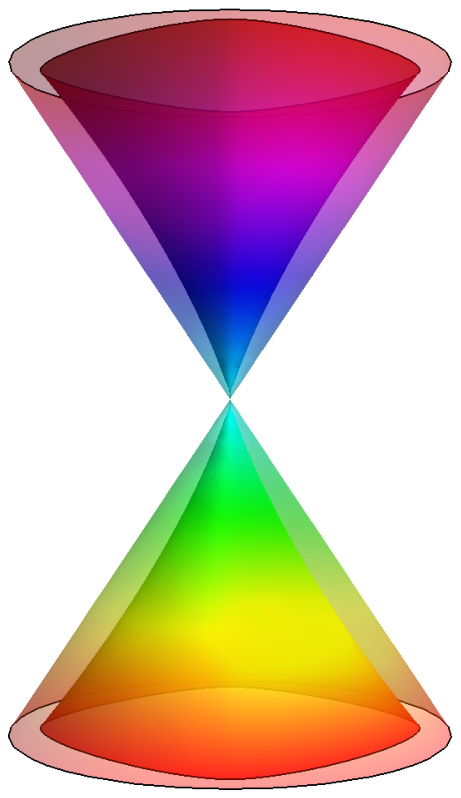}}
\caption{(a) Schematic illustration of the dispersion of 2D
semi-Dirac fermions with (inside) and without (outside) including
the effects of Coulomb interaction. (b) Renormalized (inside) and
unrenormalized (outside) dispersion 2D Dirac fermions (see
Ref.~\cite{Elias11} for more details).} \label{semi-Dirac}
\end{figure}

An interesting problem is to unambiguously determine the impact of
Coulomb interaction on the low-energy properties of semi-Dirac
fermions. For this purpose, it is necessary to go beyond both the
weak-coupling expansion and the $1/N$ expansion and find a suitable
method that is valid for any value of $\alpha$ and any value of $N$.
Recently, an efficient non-perturbative quantum-field-theoretical
approach has been developed \cite{Liu21, Pan21} to handle strong
fermion-boson couplings. The crucial procedure of this approach is
to derive the exact and self-closed Dyson-Schwinger (DS) equation of
the full fermion propagator. The contributions of the fermion-boson
vertex corrections are entirely determined by solving a number of
exact identities satisfied by various correlation functions
\cite{Liu21, Pan21}. This approach has previously been adopted to
deal with the electron-phonon interaction in ordinary metals
\cite{Liu21} and the Coulomb interaction between Dirac fermions in
2D DSMs \cite{Pan21}. Here, we apply this approach to study the
Coulomb interaction in semi-DSM. We first obtain the exact
self-consistent integral equations of the renormalization functions
for $\upsilon$ and $B$, which are valid for arbitrary values of
$\alpha$ and $N$, and then numerically solve these equations. We
find that the dispersion of semi-Dirac fermions are substantially
renormalized by the Coulomb interaction. At low energies, the
renormalized fermion dispersion is linear in momentum in two
directions, qualitatively consistent with the weak-coupling results
\cite{Kotov21}. As shown in Fig.~\ref{semi-Dirac}(a), the fermion
dispersion is strongly reshaped and becomes a Dirac cone.

The rest of the paper is organized as follows. In
Sec.~\ref{Sec:Model}, we define the effective model of the system.
In Sec.~\ref{Sec:DSE}, we present the DS equation of fermion
propagator by using four coupled Ward-Takahashi identities (WTIs).
In Sec.~\ref{Sec:Results}, we show the numerical results and discuss
the physical implications. In Sec.~\ref{Sec:Summary}, we summarize
the main results.

\section{Model \label{Sec:Model}}

The model considered in this work describes the Coulomb interaction
between 2D semi-Dirac fermions. We now present the generic form of
the action. The partition function of the system has the following
form:
\begin{eqnarray}
Z &=& \int \mathcal{D}\psi \mathcal{D}{\bar
\psi} \mathcal{D}a_0 e^{i\int dx \mathcal{L}[\psi,{\bar\psi},a_0]}, \\
\mathcal{L}[\psi,{\bar\psi},a_0] &=& \mathcal{L}_f[\psi,{\bar\psi}]
+ \mathcal{L}_e[a_0] + \mathcal{L}_{fe}[\psi,{\bar\psi},a_0].
\end{eqnarray}
Here, $x=(t,{\bf x})$ denotes the $(1+2)$-dimensional coordinate
vector and $dx = dtd{\bf x}$. Below we define the three parts of
$\mathcal{L}[\psi,{\bar\psi},a_0]$ in order.

The the Lagrangian density of free semi-Dirac fermions is
\begin{eqnarray}
\mathcal{L}_{f}[\psi,{\bar\psi}] = \sum_{\sigma=1}^N
{\bar\psi}_\sigma(x) \left({i\gamma^{0}{\partial_t} -
\mathcal{H}_f}\right) {\psi_\sigma}(x).
\end{eqnarray}
The conjugate of spinor field $\psi$ is ${\bar\psi} =
\psi^{\dag}\gamma^0$. The flavor index is denoted by $\sigma$, which
sums from $1$ to $N$. The spinor $\psi$ has two components. The
Hamiltonian density $\mathcal{H}_f$ is
\begin{eqnarray}
\mathcal{H}_f = -i \upsilon\gamma^{1}\partial_{x} -
B\gamma^{2}\partial^{2}_{y},
\end{eqnarray}
where $\upsilon$ is the velocity along the $x$ direction and
$\frac{1}{(2B)}>0$ is the mass along the $y$ direction . The three
$2\times2$ Dirac matrices can be written as $\gamma^0=\sigma_3$,
$\gamma^1=i\sigma_1$, and $\gamma^2=i\sigma_2$.

The pure Coulomb interaction is modeled by a direct density-density
coupling term
\begin{eqnarray}
H_{C} = \frac{1}{4\pi}\frac{e^{2}}{\upsilon\epsilon}\sum_{\sigma,\sigma'}
\int d^2\mathbf{x} d^2 \mathbf{x}'\rho_{\sigma}(\mathbf{x})
\frac{1}{\left|\mathbf{x} - \mathbf{x}'\right|}
\rho_{\sigma'}^{\dag}(\mathbf{x}'),
\end{eqnarray}
where the fermion density operator is $\rho_{\sigma}(\mathbf{x})
\equiv \psi_{\sigma}^{\dag}(\mathbf{x})\psi_{\sigma}(\mathbf{x}) =
{\bar \psi}_{\sigma}(\mathbf{x})\gamma^{0}
\psi_{\sigma}(\mathbf{x})$. The DS equation approach was designed to
treat fermion-boson couplings \cite{Liu21, Pan21}. In order to use
this approach, it is convenient to introduce an auxiliary scalar
field $a_{0}$ and then to re-express the Coulomb interaction by the
following two terms \cite{Pan21}
\begin{eqnarray}
\mathcal{L}_{e}[a_{0}] &=& a_{0} \frac{\mathbb{D}}{2}a_{0}, \\
\mathcal{L}_{fe}[\psi,{\bar\psi},a_{0}] &=& \sum^N_{\sigma=1}
a_{0}{\bar \psi}_{\sigma} \gamma^{0}\psi_{\sigma}.
\end{eqnarray}
After performing Fourier transformation, the inverse of the operator
$\mathbb{D}$ is converted into the free boson propagator
$D_{0}(\mathbf{q})$, which will be given later. Notice the absence
of self-coupling terms of the bosonic field $a_{0}$, which is owing
to the important fact that the Coulomb interaction originates from
an Abelian U(1) gauge symmetry.

The strength of Coulomb interaction is characterized by a
dimensionless parameter
\begin{eqnarray}
\alpha = \frac{e^{2}}{\upsilon \varepsilon_{D}},
\end{eqnarray}
where $\varepsilon_{D}$ is the dielectric constant, which can be
regarded as an effective fine structure constant. The velocity
$\upsilon$ is explicitly written down throughout this paper.

%The above Lagrangian density respects a discrete parity symmetry
%\begin{eqnarray}
%P\psi_\sigma(x,y,t)P^{-1}=\sigma_1\psi_\sigma(-x,y,t). \label{h0}
%\end{eqnarray}
%This symmetry would be dynamically broken if the interaction
%generates a finite mass term $\bar{\psi}\psi$.

\section{Dyson-Schwinger equations \label{Sec:DSE}}

In this section we present a number of DS equations and exact
identities satisfied by various correlation functions. The free
boson propagator, expressed in terms of $\alpha$, has the form
\begin{eqnarray}
D_{0}(\mathbf{q}) = \frac{2\pi \alpha\upsilon}{|\mathbf{q}|}.
\end{eqnarray}
The free fermion propagator is
\begin{eqnarray}
G_{0}(p) \equiv G_{0}(p_{0},\mathbf{p}) = \frac{1}{\gamma^0p_0 -
\upsilon\gamma^1 p_x - B\gamma^2 p^2_y}.
\end{eqnarray}
After including the interaction-induced corrections, it is
significantly renormalized and becomes
\begin{eqnarray}
\begin{aligned}
&G(p) \equiv G(p_{0},\mathbf{p}) \\
&=\frac{1}{A_0(p)\gamma^0p_0 - A_1(p) \upsilon\gamma^1 p_x -
A_2(p)B\gamma^2p^2_y},
\end{aligned}
\label{Eq:fullGp}
\end{eqnarray}
where the renormalization function $A_0(p) \equiv
A_{0}(p_{0},\mathbf{p})$ embodies the (Landau-type) fermion damping,
$A_1(p)\equiv A_{1}(p_{0},\mathbf{p})$ reflects the renormalization
of fermion velocity along the $x$-axis, and $A_2(p)\equiv
A_{2}(p_{0},\mathbf{p})$ contains the renormalization of fermion
mass along the $y$-axis.

The free and full propagators are related by the following
self-consistent DS integral equations
\begin{eqnarray}
G^{-1}(p) &=& G^{-1}_{0}(p) + i \int \frac{d^{3}k}{(2
\pi)^{3}}\gamma^{0}G(k)D(k-p)\Gamma_{\mathrm{int}}(k,p),
\label{eq:DSEG} \nonumber \\
D^{-1}(q) &=& D^{-1}_{0}(q) - i N\int\frac{d^{3}k}{(2
\pi)^{3}}{\mathrm{Tr}}\big[\gamma^{0} G(k+q) \nonumber \\
&& \times \Gamma_{\mathrm{int}}(k+q,k)G(k)\big]. \label{eq:DSED}
\end{eqnarray}
Here, $D(q)$ denotes the full boson propagator. For notational
simplicity, the DS equations are expressed in the momentum space.
These two DS equations can be derived rigorously by performing
field-theoretic analysis within the framework of functional integral
\cite{Pan21}. Here, $\Gamma_{\mathrm{int}}(k,p)$ stands for the
proper (external-legs truncated) fermion-boson vertex function
defined via the following three-point correlation function
\begin{eqnarray}
D(k-p)G(k)\Gamma_{\mathrm{int}}(k,p)G(p) = \langle \phi \psi {\bar
\psi}\rangle.\label{eq:Gammaint}
\end{eqnarray}
To determine $G(p)$ and $D(q)$, one needs to first specify the
vertex function $\Gamma_{\mathrm{int}}(k,p)$. By carrying out
functional calculations, one can show that $\Gamma_{\mathrm{int}}$
satisfies its own DS equation
\begin{eqnarray}
\Gamma_{\mathrm{int}}(k,p) &=& -\gamma^{0} + \int \frac{d^{3}p'}{(2
\pi)^{(3)}}G(p'+k) \Gamma_{\mathrm{int}}(k,p') \nonumber \\
&& \times G(p')K_{4}(p,p',k),
\end{eqnarray}
where $K_{4}(p,p',q)$ represents a kernel function that is related
to the four-point correlation function $\langle \psi{\bar
\psi}\psi{\bar \psi}\rangle$ as follows
\begin{eqnarray}
G(p+p'+k)G(p')K_{4}(p,p',k)G(p)G(k) = \langle \psi {\bar \psi}\psi
{\bar \psi}\rangle.
\end{eqnarray}
The function $K_{4}(p,p',q)$ also satisfies a peculiar DS integral
equation, which in turn is linked to five-, six-, and higher-point
correlation functions. Repeating such manipulations, one would
obtain an infinite hierarchy of coupled integral equations
\cite{Liu21, Pan21}. The full set of DS integral equations are exact
and can give us all of the interaction-induced effects. However,
such equations seem to be of little use in practice since they are
not closed and cannot be really solved.

To make such DS equations solvable, one needs to find a proper way
to introduce truncations. The strategy of the widely used
Migdal-Eliashberg (ME) theory is to simply discard all the vertex
corrections by replacing the full vertex function with the bare one,
i.e.,
\begin{eqnarray}
\Gamma_{\mathrm{int}}(k,p) \to -\gamma^{0}.
\end{eqnarray}
Then the originally infinite number of DS equations are reduced to
only two equations of $G(p)$ and $D(q)$, namely
\begin{eqnarray}
G^{-1}(p) &=& G^{-1}_{0}(p) - i \int \frac{d^{3}k}{(2
\pi)^{3}}\gamma^{0}G(k)D(k-p)\gamma^{0},
\label{eq:MEEG} \nonumber \\
D^{-1}(q) &=& D^{-1}_{0}(q) + i N\int\frac{d^{3}k}{(2\pi)^{3}}
{\mathrm{Tr}}\big[\gamma^{0}G(k+q)\gamma^{0}G(k)\big].
\nonumber
\label{eq:MEED}
\end{eqnarray}
These two equations are self-closed and solvable. However, the
validity of the ME theory is indeed unjustified, especially in the
strong-coupling regime.

The non-perturbative approach developed in Refs.~\cite{Liu21, Pan21}
aims to take into account the contributions of all the vertex
function $\Gamma_{\mathrm{int}}(k,p)$ by properly using several
exact identities obeyed by some correlation functions. Below we
briefly outline how this approach works in the case of 2D semi-DSM.
More details about this approach can be found in Refs.~\cite{Liu21,
Pan21}.

Now let us introduce a generic current operator
\begin{eqnarray}
j^\mu_M(x)=\bar{\psi}_\sigma(x) M^\mu \psi_\sigma(x)
\end{eqnarray}
based on four matrices
\begin{eqnarray}
M^\mu=\left(\gamma^0, \gamma^1, \gamma^2, \gamma^{012}\right),
\end{eqnarray}
where $\gamma^{012}=\gamma^{0}\gamma^{1}\gamma^{2}$. This current
can be used to define a corresponding current vertex function
$\Gamma^\mu_M$ as follows
\begin{eqnarray}
&&\langle j^\mu_M(x)\psi_\alpha(y)\bar{\psi}_\beta(z) \rangle
\nonumber \\
&=& \int d\xi_1 d\xi_2 \left(G(y-\xi_1)\Gamma^\mu_M(\xi_1 - x,
x-\xi_2) G(\xi_2-z)\right)_{\alpha\beta}. \label{h0}\nonumber \\
\end{eqnarray}
The four components of the current vertex function $\Gamma^\mu_M$
satisfy four self-consistent generalized WTIs. Solving these WTIs,
one can express the current vertex functions as a linear combination
of the inverse of the full fermion propagator $G(p)$. The procedure
of deriving such WTIs is illustrated in great detail in
Ref.~\cite{Pan21} and will not be repeated here. Below we only
present the WITs that relate $\Gamma^\mu_M$ to $G(p)$.

We use $\Gamma_{\gamma^{0}}$, $\Gamma_{\gamma^{1}}$,
$\Gamma_{\gamma^{2}}$, and $\Gamma_{\gamma^{012}}$ to denote the
four components of $\Gamma^\mu_M$. For notational simplicity, let us
define several quantities here: $q_0=k_0-p_0$, $P_0=k_0+p_0$,
$q_1=k_1-p_1$, $P_1=k_1+p_1$, $q_2=k_2-p_2$, and $P_2=k_2+p_2$.
According to the detailed analysis presented in Ref.~\cite{Pan21},
these four current vertex functions and the full fermion propagator
are linked to each other via the following four WTIs
\begin{eqnarray}
\mathbf{M}_{\mathcal{A}}\begin{pmatrix}\Gamma_{\gamma^0}\\
\Gamma_{\gamma^1}\\ \Gamma_{\gamma^2} \\
\Gamma_{\gamma^{012}}\\\end{pmatrix} =
\begin{pmatrix}\mathcal{A}_0\\ \mathcal{A}_1 \\ \mathcal{A}_2 \\
\mathcal{A}_3 \\
\end{pmatrix},
\end{eqnarray}
where the matrix
\begin{eqnarray}
\mathbf{M}_{\mathcal{A}}=\begin{pmatrix}q_0 & q_1 & -q_2 & 0\\ -q_1
& -q_0 & 0 & P_2 \\ q_2 & 0 & -q_0 & P_1\\ 0 & -P_2 & -P_1 &
q_0\\\end{pmatrix}, \label{eq:matrixm}
\end{eqnarray}
and
\begin{eqnarray}
\mathcal{A}_0 &=& - G^{-1}(k)+G^{-1}(p),\label{eq:a0}\\
\mathcal{A}_1 &=& G^{-1}(k)\gamma^{0}\gamma^{1}
+\gamma^{0}\gamma^{1}G^{-1}(p), \label{eq:a1} \\
\mathcal{A}_2 &=& G^{-1}(k)\gamma^{0}\gamma^{2}
+\gamma^{0}\gamma^{2}G^{-1}(p),\label{eq:a2} \\
\mathcal{A}_3 &=& - G^{-1}(k)\gamma^{1}\gamma^{2} +
\gamma^{1}\gamma^{2}G^{-1}(p).\label{eq:a3}
\end{eqnarray}
After solving the above four coupled identities, each of the four
current vertex functions, namely $\Gamma_{\gamma^{0}}$,
$\Gamma_{\gamma^{1}}$, $\Gamma_{\gamma^{2}}$, and
$\Gamma_{\gamma^{012}}$, can be determined and expressed purely in
terms of the fermion propagator $G(p)$. Notice that the matrix
$\mathbf{M}_{\mathcal{A}}$ given by Eq.~(\ref{eq:matrixm}) is
different from that obtained in the case of 2D DSM \cite{Pan21} due
to the difference in the fermion dispersions.

Apparently, the current vertex functions do not enter into the
original DS equations of $G(p)$ and $D(q)$. To make the above WTIs
useful, we should find a way to substitute the current vertex
functions into the DS equation of $G(p)$. As demonstrated previously
in Ref.~\cite{Pan21}, there exists an identity that connects
$\Gamma_{\gamma^0}(k, p)$ to $D_0(k-p)$, $D(k-p)$, and
$\Gamma_{\mathrm{int}}(k, p)$. Such an identity also exists in the
case of 2D semi-DSM, given by
\begin{eqnarray}
D_0(k-p)\Gamma_{\gamma^0}(k,p) = D(k-p) \Gamma_{\mathrm{int}}(k,p).
\label{identity5}
\end{eqnarray}
It is now clear that we only need one of the four current vertex
functions, i.e., $\Gamma_{\gamma^0}(k,p)$. After solving the four
coupled WTIs, $\Gamma_{\gamma^0}$ can be easily obtained:
\begin{eqnarray}
\Gamma_{\gamma^0}(k,p) &=& \frac{1}{\mathrm{det}
(\mathbf{M}_{\mathcal{A}})}\Big[q_0\left(q_0^2-P_1^2-P_2^2\right)
\mathcal{A}_0 \nonumber \\
&& -\left(q_1P_{1}^{2}+q_2P_1P_2-q_{0}^{2}q_1\right)\mathcal{A}_1
\nonumber \\
&& -\left(q_{0}^{2}q_2-q_1P_1P_2-q_2P_{2}^{2}\right)\mathcal{A}_2
\nonumber \\
&& +q_0\left(q_2P_1-q_1P_2\right)\mathcal{A}_3\Big],
\label{eq:gammagamma0}
\end{eqnarray}
where the denominator is
\begin{eqnarray}
\mathrm{det}(\mathbf{M}_{\mathcal{A}}) = q_0^4 - q^2_0\left(q^2_1 +
q^2_2 + P^2_1 + P^2_2\right) + \left(P_1q_1 + P_2q_2\right)^2.
\nonumber \\
\end{eqnarray}
According to the identity (\ref{identity5}), the product $D(k-p)
\Gamma_{\mathrm{int}}(k,p)$ appearing in the DS equation of $G(p)$
can be replaced with the product $D_0(k-p)\Gamma_{\gamma^0}(k,p)$,
which leads to
\begin{equation}
G^{-1}(p)=G_0^{-1}(p)+i\int \frac{d^{3}k}{(2\pi)^{3}}
\gamma^{0}G(k)D_{0}(k-p)\Gamma_{\gamma^{0}}(k,p), \label{eq:DSEGP}
\end{equation}
where $\Gamma_{\gamma^{0}}(k,p)$, as shown by
Eq.~(\ref{eq:gammagamma0}) and Eqs.~(\ref{eq:a0}-\ref{eq:a3}),
depends solely on the full fermion propagator. Clearly, the DS
equation of $G(p)$ becomes entirely self-closed. To investigate the
renormalization of fermion dispersion, we substitute the generic
form of $G(p)$, i.e., Eq.~(\ref{Eq:fullGp}), into
Eq.~(\ref{eq:DSEGP}) and obtain
\begin{widetext}
\begin{eqnarray}
&&A_{0}(p)\gamma^{0}p_{0}-A_1(p)\upsilon\gamma^{1}p_{x} -
A_2(p)B\gamma^{2}p^{2}_{y} - \gamma^{0}p_{0}+\upsilon\gamma^{1}p_{x}
+ B\gamma^{2}p^{2}_{y}
\nonumber \\
&=& i\int \frac{d^{3}k}{(2\pi)^3}\gamma^{0}\frac{A_0(k)\gamma^{0}
k_{0}-A_1(k)\upsilon\gamma^{1}k_{x}-A_2(k) B\gamma^{2}
k^{2}_{y}}{A^{2}_{0}(k)k^{2}_{0}-A_1^2(k)\upsilon^2 k^{2}_{x} -
A^2_2(k)B^2 k^{4}_{y}} D_0(k-p)\Gamma_{\gamma^0}(k,p).
\label{eq:EQGP}
\end{eqnarray}
\end{widetext}

This DS equation can be readily decomposed into three coupled
integral equations of $A_0(p)$, $A_1(p)$, and $A_2(p)$. Multiplying
three matrices $\gamma^0$, $\gamma^1$, and $\gamma^2$ to both sides
of Eq.~(\ref{eq:EQGP}) and then calculating the trace would lead us
to three coupled integral equations of $A_0(p)$, $A_1(p)$, and
$A_2(p)$, respectively. All the interaction-induced quantum
many-body effects of Dirac fermions can be extracted from the
numerical solutions of $A_0(p)$, $A_1(p)$, and $A_2(p)$.

The three integral equations can be solved by means of the iteration
method. The detailed procedure of implementing this method has
already been demonstrated previously in Ref.~\cite{Liu21}. These
equations contain an integration over three variables, including
$k_{0}$, $k_{x}$, and $k_{y}$. It would consume an extremely long
computational time to integrate over three variables by using the
iteration method. In order to simplify numerical computations, we
introduce an instantaneous approximation and assume that the
frequency dependence of $A_{0,1,2}(p)$ is rather weak. Throughout
this paper, we fix the functions $A_{0,1,2}(p)$ at zero energy,
i.e., $p_0 = 0$, and consider only the momentum dependence of
$A_{0,1,2}(p_{x},p_{y})$. In this approximation, it is easy to see
that $A_0=1$, we get a coupled system of equations for
$A_1(\mathbf{p})$ and $A_2(\mathbf{p})$:
\begin{widetext}
\begin{eqnarray}
A_1(\mathbf{p}) &=& 1 + \frac{1}{p_x}\int\frac{d\epsilon
d^2\mathbf{k}}{(2\pi)^3}\frac{D_0(\mathbf{k}-\mathbf{p})}{\epsilon^2
+ A_1^2(\mathbf{k})k^2_x + A^2_2(\mathbf{k})\beta^{2}k^{4}_{y}}
\frac{1}{\mathrm{det}(\mathbf{M}_{\mathcal{A}})}
\nonumber \\
&& \times \Big\{\epsilon^{2}\Big[\left(q_xP_x^2 + q_yP_xP_y +
\epsilon^{2}q_x\right) + \left(\epsilon^2+P_{x}^{2} +
P_y^2\right)\left(A_1(\mathbf{p})p_x - A_1(\mathbf{k})k_{x}\right) \nonumber \\
&& +\left(q_xP_y-q_yP_x\right)\left(A_2(\mathbf{p})\beta p^{2}_{y} +
A_2(\mathbf{k})\beta k^{2}_{y}\right)\Big]\nonumber \\
&& +A_1(\mathbf{k})k_x\Big[\epsilon^{2}\left(\epsilon^{2} +
P_{x}^{2} + P_{y}^{2}\right)-\left(q_x P_{x}^{2} + q_{y}P_{x}P_{y} +
\epsilon^{2}q_{x}\right)\left(A_1(\mathbf{p}) p_{x} -
A_{1}(\mathbf{k})k_x\right) \nonumber \\
&& -\left(q_{y}P_{y}^{2} + q_xP_x P_y + \epsilon^{2}q_{y}\right)
\left(A_2(\mathbf{p})\beta p^{2}_{y}- A_2(\mathbf{k})\beta
k^{2}_{y}\right)\Big] \nonumber \\
&& -A_2(\mathbf{k})\beta k^2_y\Big[\epsilon^2 \left(q_x P_y-
q_yP_x\right) + \left(q_{y}P_{y}^{2}+q_{x}P_{x}P_{y} +
\epsilon^{2}q_{y}\right)\left(A_1(\mathbf{p})p_{x} +
A_1(\mathbf{k}) k_{x}\right) \nonumber \\
&& -\left(q_{x}P_{x}^{2}+q_{y}P_{x}P_{y}+\epsilon^{2}q_{x}\right)
\left( A_2(\mathbf{p})\beta p^2_y+A_2(\mathbf{k}) \beta
k^{2}_{y}\right)\Big]\Big\}, \label{A1eq} \\
A_2(\mathbf{p}) &=& 1 + \frac{1}{\beta p^2_y}\int\frac{d\epsilon
d^2\mathbf{k}}{(2\pi)^3}\frac{D_0(\mathbf{k}-\mathbf{p})}{\epsilon^2
+ A_1^2(\mathbf{k})k^2_x + A^2_2(\mathbf{k})\beta^2k^4_y}
\frac{1}{\mathrm{det}(\mathbf{M}_{\mathcal{A}})}
\nonumber \\
&&\times\Big\{\epsilon^2\Big[\left(q_yP_y^2+q_xP_xP_y +
\epsilon^2q_y\right)-\left(q_xP_y- q_yP_x\right)
\left(A_1(\mathbf{p})p_x + A_1(\mathbf{k}) k_x\right) \nonumber \\
&& +\left(\epsilon^2+P_x^2+P_y^2\right)\left(A_2(\mathbf{p}) \beta
p^{2}_{y} - A_2(\mathbf{k})\beta k^2_y\right)\Big]\nonumber \\
&& +A_1(\mathbf{k}) k_x\Big[\epsilon^2\left(q_xP_y - q_yP_x \right)
+ \left(q_yP_y^2+q_xP_xP_y+\epsilon^2q_y\right)\left(A_1(\mathbf{p})
p_x + A_1(\mathbf{k})k_x\right) \nonumber \\
&&-\left(q_xP_x^2 + q_yP_x P_y + \epsilon^{2}q_{x}\right)
\left(A_2(\mathbf{p})\beta p^{2}_{y} + A_2(\mathbf{k}) \beta
k^{2}_{y}\right)\Big]\nonumber\\
&& +A_2(\mathbf{k})\beta k^{2}_{y} \Big[\epsilon^2\left(\epsilon^2 +
P_{x}^{2} + P_{y}^{2}\right) - \left(q_xP_{x}^{2}+q_{y}P_{x}P_{y}+
\epsilon^{2}q_{x}\right)\left(A_1(\mathbf{p})p_{x} -
A_{1}(\mathbf{k})k_{x}\right) \nonumber \\
&& -\left(q_{y}P_{y}^{2}+q_{x}P_{x}P_{y} + \epsilon^{2}q_{y}\right)
\left(A_2(\mathbf{p})\beta p^{2}_{y} - A_2(\mathbf{k})\beta
k^{2}_{y} \right)\Big]\Big\}. \label{A2eq}
\end{eqnarray}
\end{widetext}
The integration ranges for $k_x$ and $k_y$ are chosen as $k_x \in
(-\Lambda_{x},\Lambda_{x})$ and $k_y \in(-\Lambda_{y},\Lambda_{y})$,
respectively. In practise, it is sufficient to take $\Lambda_{x} =
\Lambda_{y} = \Lambda$, where $\Lambda$ is an UV cutoff of momentum.
In principle, the energy $\epsilon$ takes all the possible values,
namely $\epsilon \in (-\infty,\infty)$. In practical numerical
computations, the energy is integrated within the range of
$(-\Lambda_{\epsilon},\Lambda_{\epsilon})$, where the cutoff
$\Lambda_{\epsilon}$ should be taken to be sufficiently large to
make sure that the results are nearly independent of
$\Lambda_{\epsilon}$. It is convenient to re-scale all the momenta
as follows:
\begin{eqnarray}
k_x\rightarrow k_x/\Lambda, \quad k_y\rightarrow k_y/\Lambda.
\end{eqnarray}
Then the integration range is altered to $k_{x,y} \in(-1, 1)$. After
dividing the left and right sides of the original integral equations
by $\upsilon$, the integration range of energy is re-scaled to
$\epsilon \in\left(-\frac{\Lambda_\epsilon}{\upsilon\Lambda},
\frac{\Lambda_\epsilon}{\upsilon\Lambda}\right)$ with
$\upsilon\Lambda$ being the unit of energy. In our numerical
calculations, we choose $\epsilon \in(-10,10)$. The model contains
three tuning parameters: the interaction strength $\alpha$, the
velocity $\upsilon$, and a tuning parameter
$\beta=B\Lambda/\upsilon$. In the above two equations, we have
$D_0(\mathbf{k}-\mathbf{p}) = \frac{2\pi\alpha}
{\sqrt{(k_x-p_x)^2+(k_y- p_y)^2}}$, $q_x=k_x-p_x$, $P_x=k_x+p_x$,
$q_y=\beta k^2_y-\beta p^2_y$, $P_y=\beta k^2_y+\beta p^2_y$, and
$\mathrm{det}(\mathbf{M}_{\mathcal{A}})=\epsilon^4+2\epsilon^2(
k_x^2+p_x^2+\beta^2k^4_y+\beta^2p^4_y) +
(k^2_x-p^2_x+\beta^2k^4_y-\beta^2p^4_y)^2$.

\section{Renormalization of fermion dispersion \label{Sec:Results}}

In the non-interacting limit, the fermion spectrum exhibits a linear
dependence on $p_{x}$ with a coefficient $\upsilon$ and a quadratic
dependence on $p_{y}$ with a coefficient $B$. When the correlation
effects are incorporated, these two coefficients $\upsilon$ and $B$
will be renormalized, described by the functions
$A_{1}(p_{x},p_{y})$ and $A_{2}(p_{x},p_{y})$, respectively. The
interaction-induced modification of fermion dispersion should be
extracted from the numerical solutions of $A_{1}(p_{x},p_{y})$ and
$A_{2}(p_{x},p_{y})$. Throughout this section, the momentum (such as
$p_{x}$ or $p_{y}$) is in unit of $\Lambda$ and the energy is in
unit of $\upsilon \Lambda$.

\begin{widetext}

\begin{figure}[H]
\centering
\includegraphics[width=2.2in]{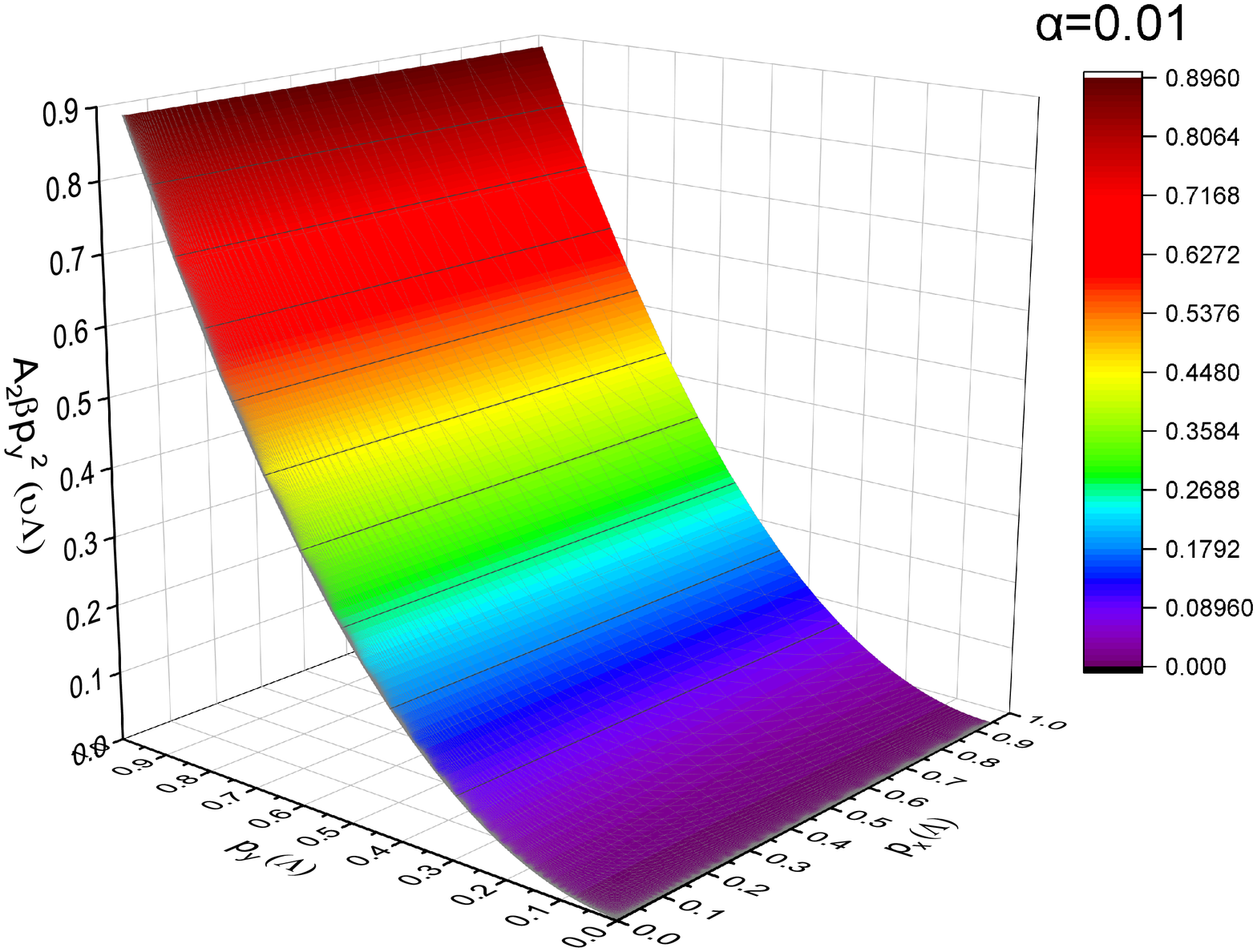}
\includegraphics[width=2.2in]{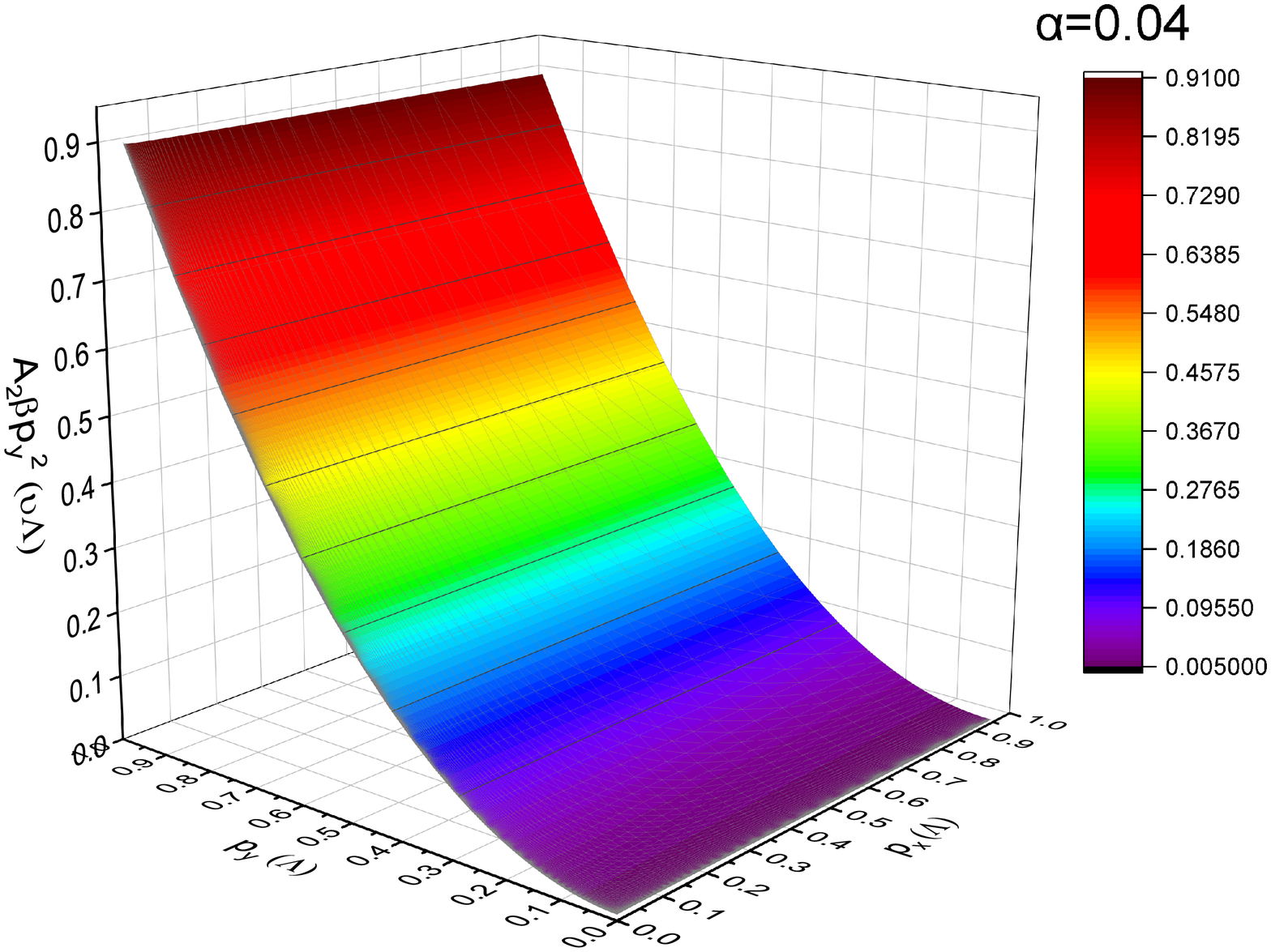}
\includegraphics[width=2.2in]{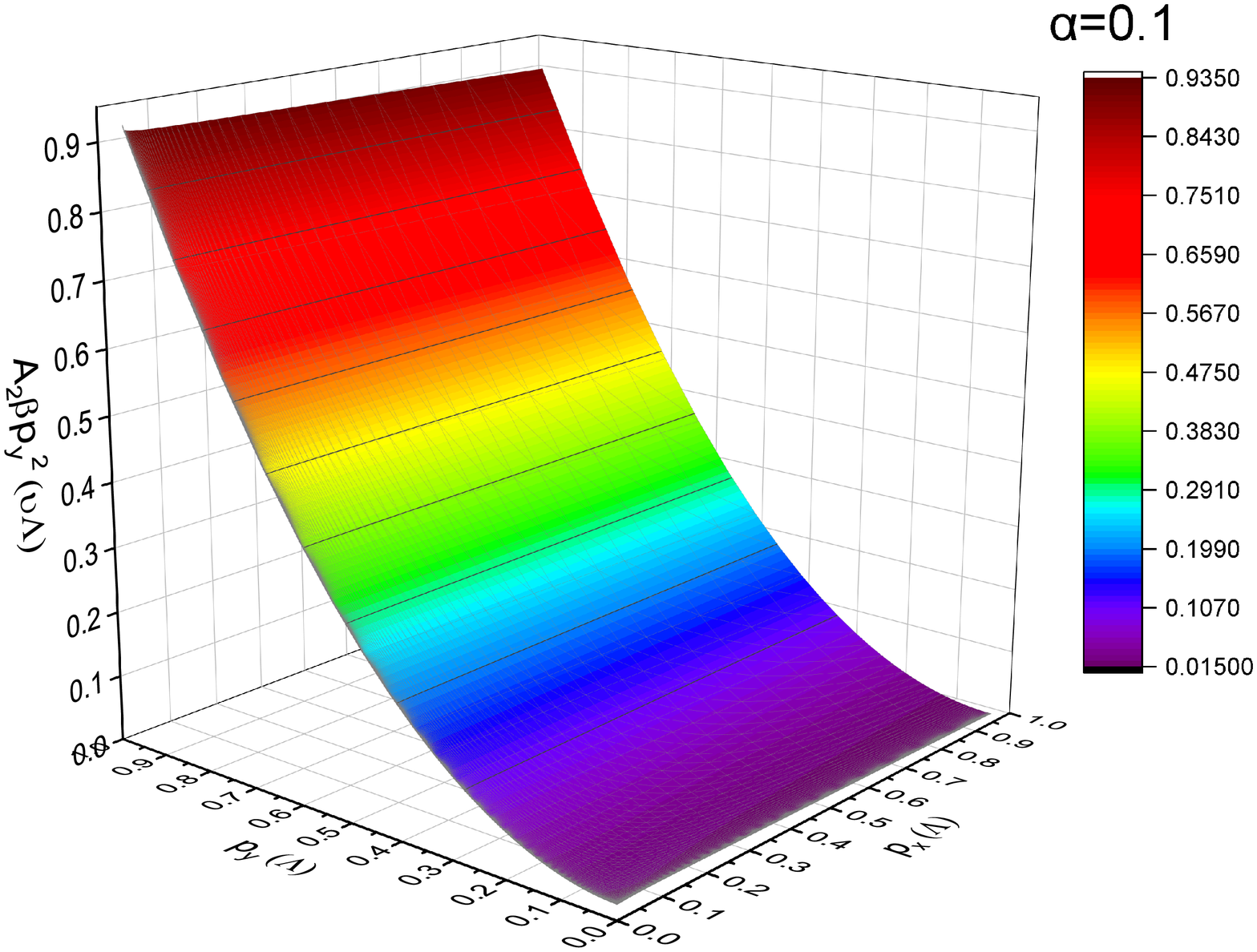}
\includegraphics[width=2.2in]{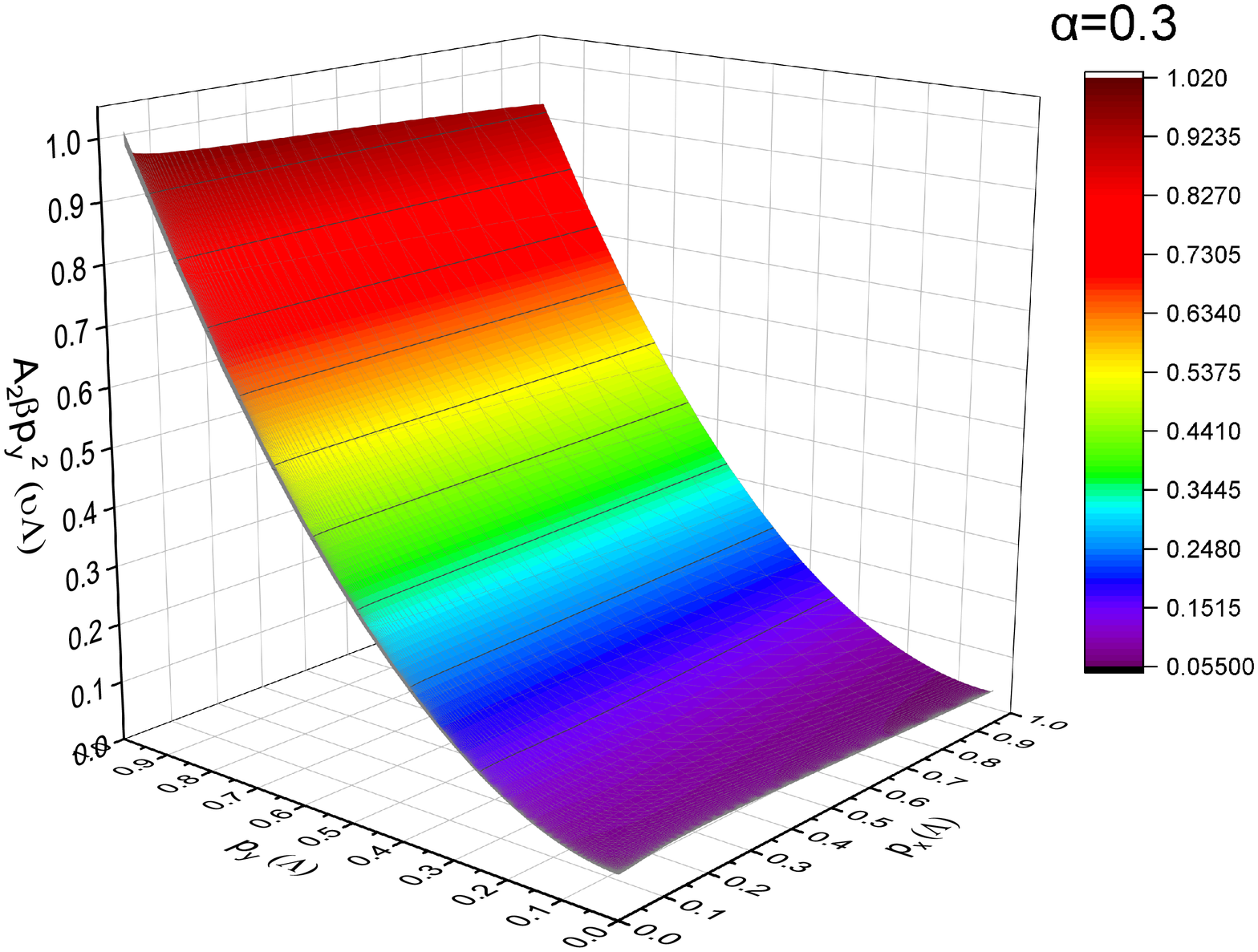}
\includegraphics[width=2.2in]{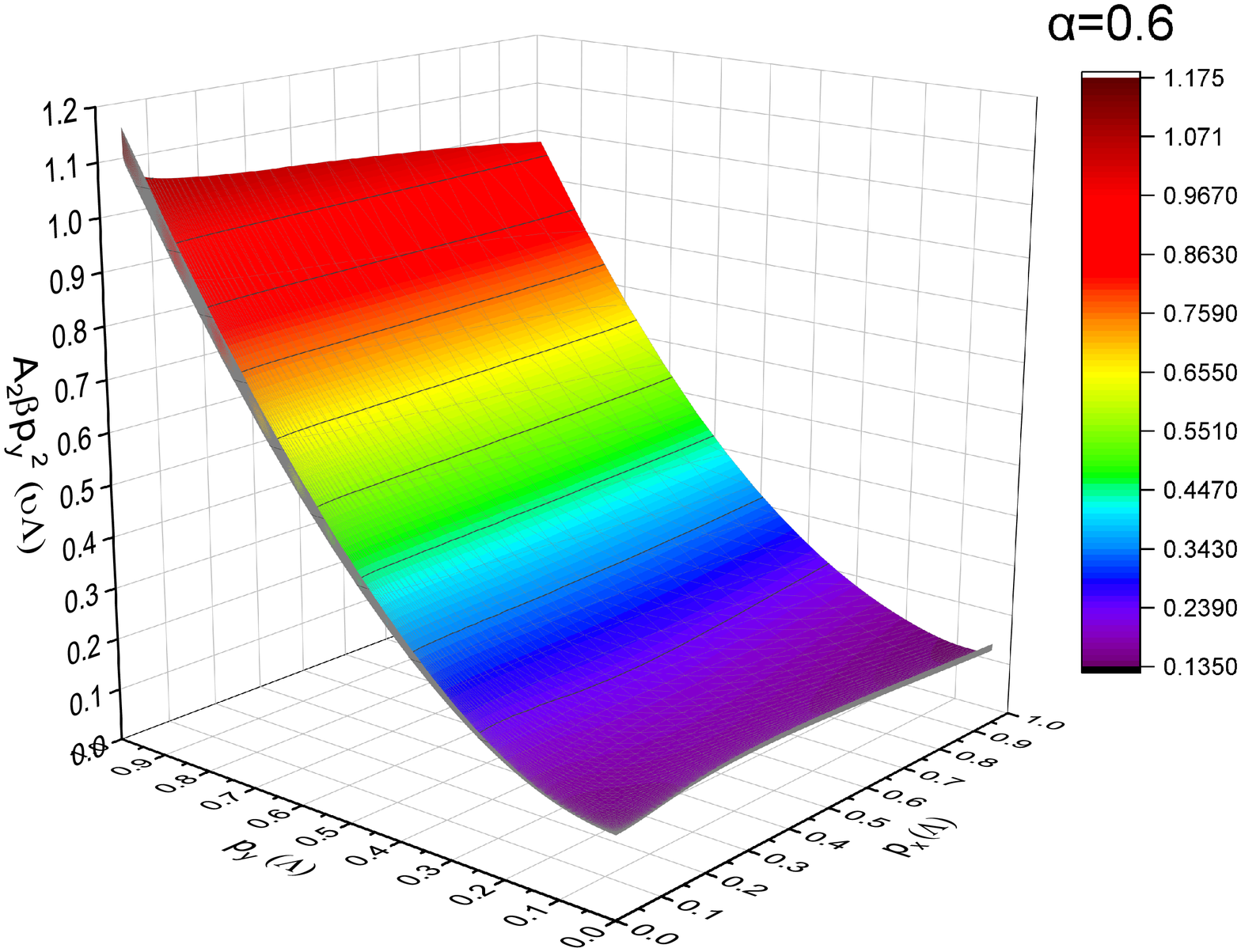}
\includegraphics[width=2.2in]{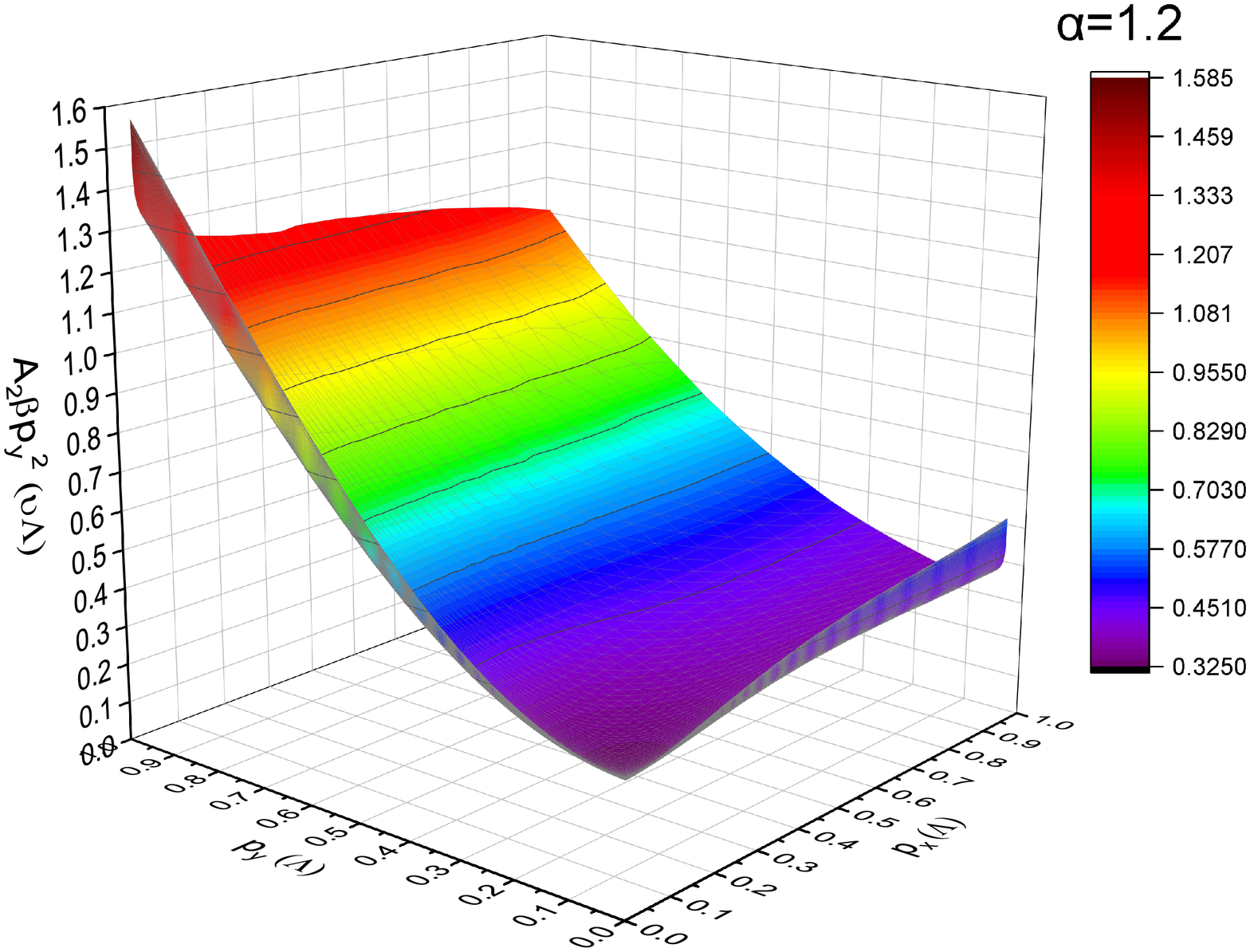}
\caption{The full momentum dependence of the function $A_{2}(p_x,
p_y)\beta p^2_y$ obtained by solving the self-consistent integral
equations of $A_{1}(\mathbf{p})$ and $A_{2}(\mathbf{p})$. We choose
six different values of $\alpha$, including $\alpha=0.01$,
$\alpha=0.04$, $\alpha=0.1$, $\alpha=0.3$, $\alpha=0.6$ and
$\alpha=1.2$. The parameter $\beta$ is taken as $\beta=1.0$. Over a
wide range of $p_x$ and $p_y$, $A_{2}(p_x, p_y)\beta p^2_y$ exhibits
a linear dependence on $p_y$ but is nearly independent of $p_x$.
Close to the small momentum $p_y$, $A_{2}(p_x, p_y)\beta p^2_y$
appears to deviate from the linear behavior, which stems from the
opening of a finite gap. Such a deviation would disappear once this
gap is closed by carefully tuning the system to approach the
semi-DSM QCP.} \label{revf3d}
\end{figure}

\end{widetext}

Let us first discuss the behavior of renormalization function
$A_2(p_x,p_y)$. We find it more convenient to consider the momentum
dependence of the function $A_2(p_x,p_y)\beta p^{2}_{y}$, instead of
$A_2(p_x,p_y)$ itself. The full momentum dependence of
$A_{2}(p_x,p_y)\beta p_{y}^{2}$ obtained by solving
Eqs.~(\ref{A1eq}-\ref{A2eq}) in Fig.~\ref{revf3d}, for six different
values of $\alpha$. It is easy to find that $A_{2}(p_x,p_y)\beta
p_{y}^{2}$ is nearly independent of $p_x$ for small values of
$\alpha$. As $\alpha$ exceeds $0.6$, $A_{2}(p_x,p_y)\beta p_{y}^{2}$
exhibits a considerable dependence on $p_{x}$. In contrast,
$A_{2}(p_x,p_y)\beta p_{y}^{2}$ manifests a significant dependence
on $p_y$ for all values of $\alpha$. To show these properties more
explicitly, we plot the $p_{x}$-dependence of $A_{2}(p_x,p_y=0)$ in
Fig.~\ref{A2pxpy}(a) and the $p_{y}$-dependence of
$A_{2}(p_x=0,p_y)$ in Fig.~\ref{A2pxpy}(b).

The $p_{y}$-dependence of $A_{2}(p_x,p_y)\beta p_{y}^{2}$ deserves a
little more analysis. Over a wide range of $p_{y}$ below the UV
cutoff, $A_{2}(p_x,p_y)\beta p_{y}^{2}$ displays a linear dependence
on $p_{y}$. This behavior indicates that the originally quadratic
fermion dispersion along the $y$-direction is turned into a linear
dispersion owing to the Coulomb interaction. Consequently, the 2D
semi-Dirac fermions behave in the same way as the ordinary 2D Dirac
fermions that exhibit a linear dispersion along two directions. For
small values of $p_{y}$, the dispersion is no longer linear in
$p_{y}$. Indeed, a finite energy gap $\Delta_{0}$ is opened at
$p_{y}=0$ and such a gap is an increasing function of $\alpha$. The
interaction-induced quadratic-to-linear transition of the fermion
dispersion along $y$-direction and also the generation of a finite
gap have already been obtained previously in the weak-coupling
perturbative calculations of Kotov \emph{et al.}. \cite{Kotov21}.
Our non-perturbative studies show that these two conclusions remain
correct even when the interaction parameter $\alpha$ becomes large.

Summarizing the results shown in Fig.~\ref{revf3d} and
Fig.~\ref{A2pxpy}(b), we find that $A_{2}(p_x,p_y)$ can be
approximately described by
\begin{eqnarray}
A_{2}(p_x, p_y)\sim \frac{a}{p_y}+\frac{b}{p_y^2},
\label{eq:A2expression}
\end{eqnarray}
where $a$ and $b$ are two fitting constants. We assume that $a \gg
b$ in the limit of $p_y \rightarrow \Lambda$ and that $a \ll b$ in
the limit of $p_y \rightarrow 0$. For large values of $p_{y}$, the
first term dominates over the second term, leading to
\begin{eqnarray}
A_{2}(p_x,p_y)\beta p_{y}^{2} \sim p_{y}.
\end{eqnarray}
For small values of $p_{y}$, the second term dominates such that
\begin{eqnarray}
A_{2}(p_x,p_y)\beta p_{y}^{2} \sim b\beta,
\end{eqnarray}
where the constant $b\beta$ represents a finite gap. Notice that
such a gap does not break any symmetry of the system and persists
for an arbitrary weak Coulomb interaction. Its existence implies
that the semi-DSM state is unstable against Coulomb interaction and
can be easily turned into a BI. However, one can tune this gap to
vanish by carefully varying certain external parameters
\cite{Kotov21}, which then would drive the system to approach the
DSM-to-BI QCP. In this regard, the second term of $A_{2}(p_x, p_y)$
can be simply dropped, leaving us with only the first term.

\begin{widetext}

\begin{figure}[H]
\centering \subfigure[]{\label{Fig.sub.1}
\includegraphics[width=2.9in]{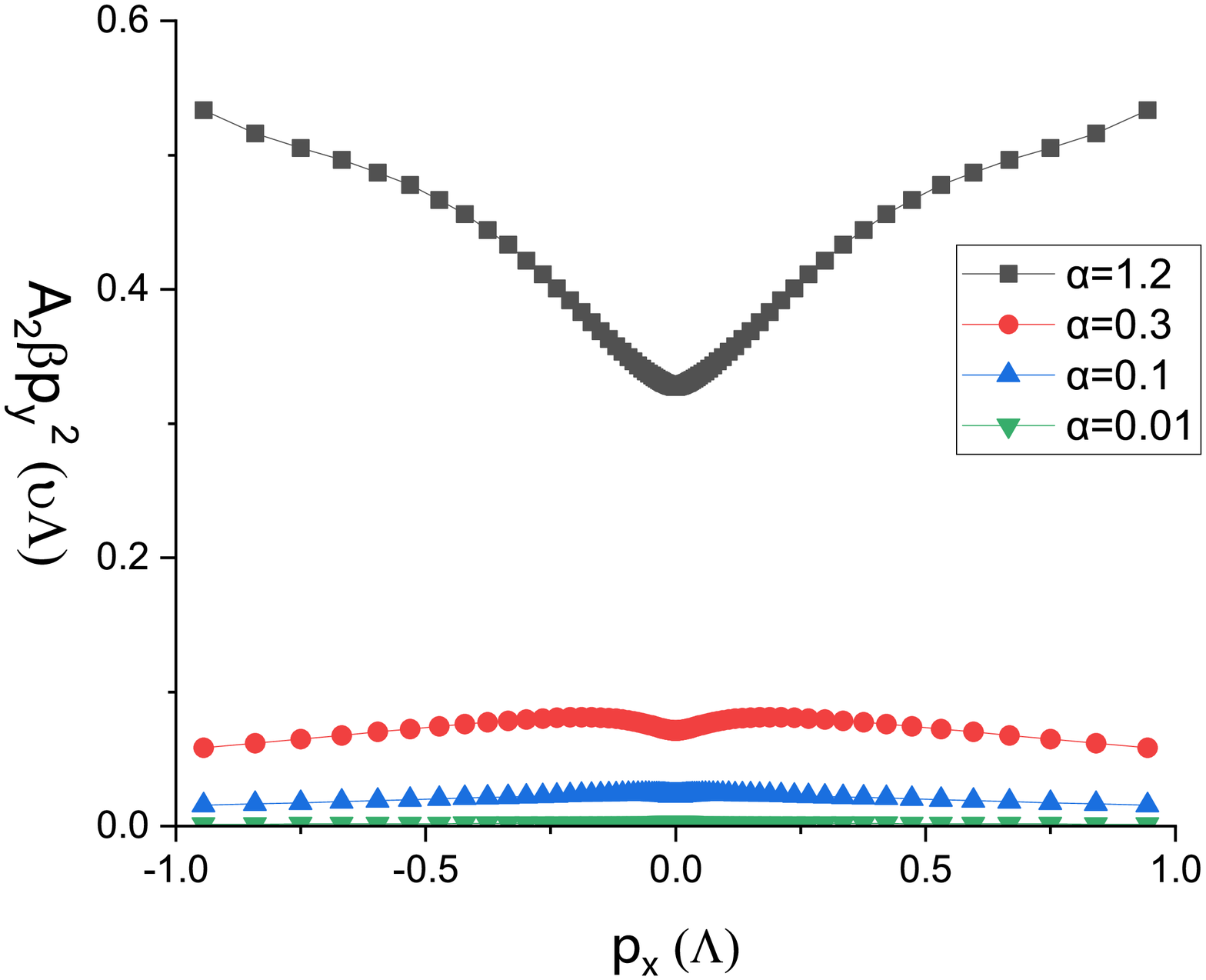}}
\subfigure[]{\label{Fig.sub.2}
\includegraphics[width=2.9in]{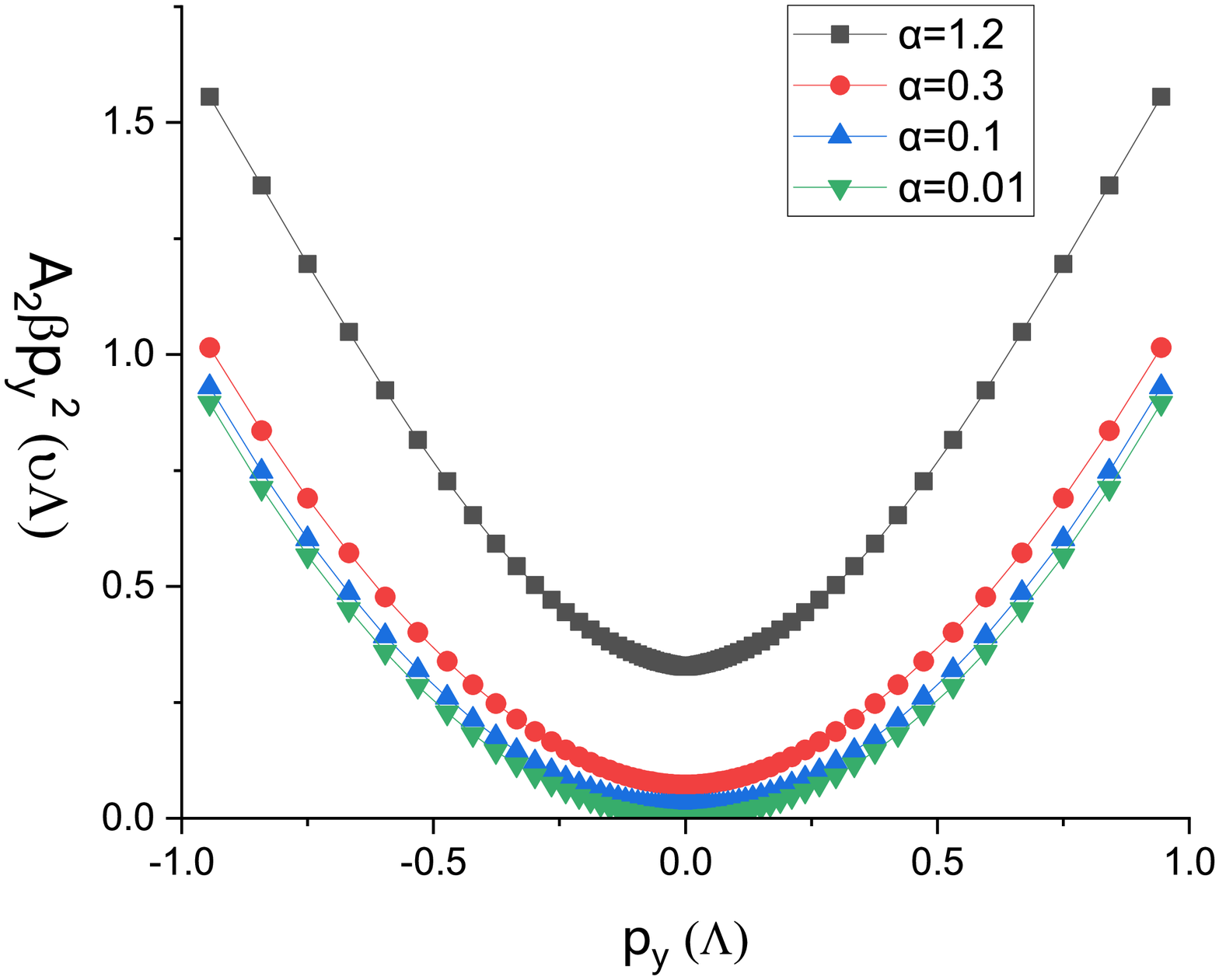}}
\subfigure[]{\label{Fig.sub.3}
\includegraphics[width=2.9in]{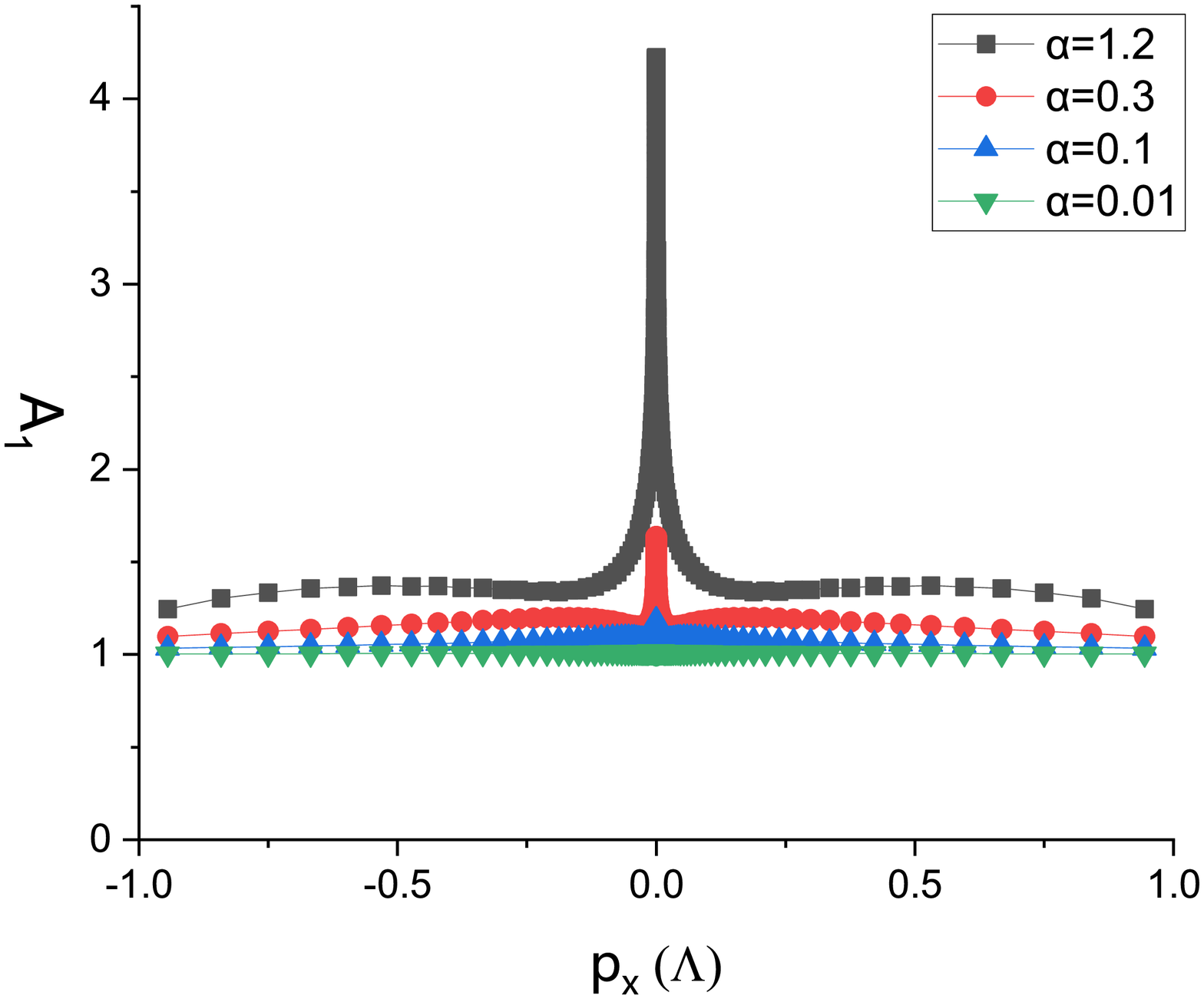}}
\subfigure[]{\label{Fig.sub.4}
\includegraphics[width=2.9in]{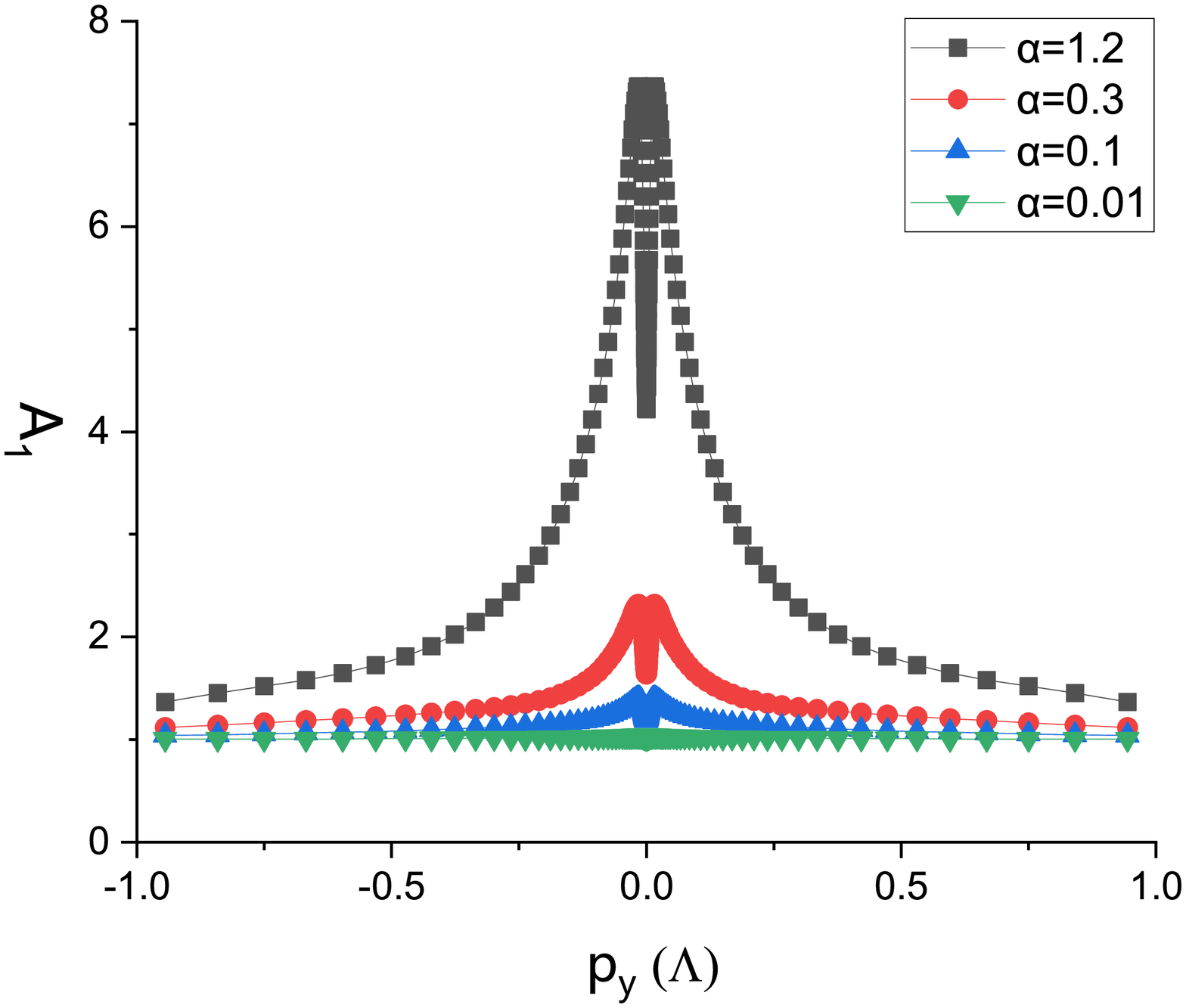}}
\caption{(a) The $p_{x}$-dependence of the function
$A_{2}(p_{x},p_{y})\beta p^{2}_{y}$ obtained in the
$p_{y}\rightarrow 0$ limit. (b) The $p_{y}$-dependence of the
function $A_{2}(p_{x},p_{y})\beta p^{2}_{y}$ obtained in the
$p_{x}\rightarrow 0$ limit. (c) The $p_{x}$-dependence of the
function $A_{1}(p_{x},p_{y})$ obtained in the $p_{y}\rightarrow 0$
limit. (d) The $p_{y}$-dependence of the function
$A_{1}(p_{x},p_{y})$ obtained in the $p_{x}\rightarrow 0$ limit.
Here, we set $\beta=1.0$ and choose four different values of
$\alpha$, including $\alpha=0.01$, $\alpha=0.1$, $\alpha=0.3$ and
$\alpha=1.2$.} \label{A2pxpy}
\end{figure}

\begin{figure}[H]
\centering \subfigure[]{\label{Fig.sub.1}
\includegraphics[width=2.9in]{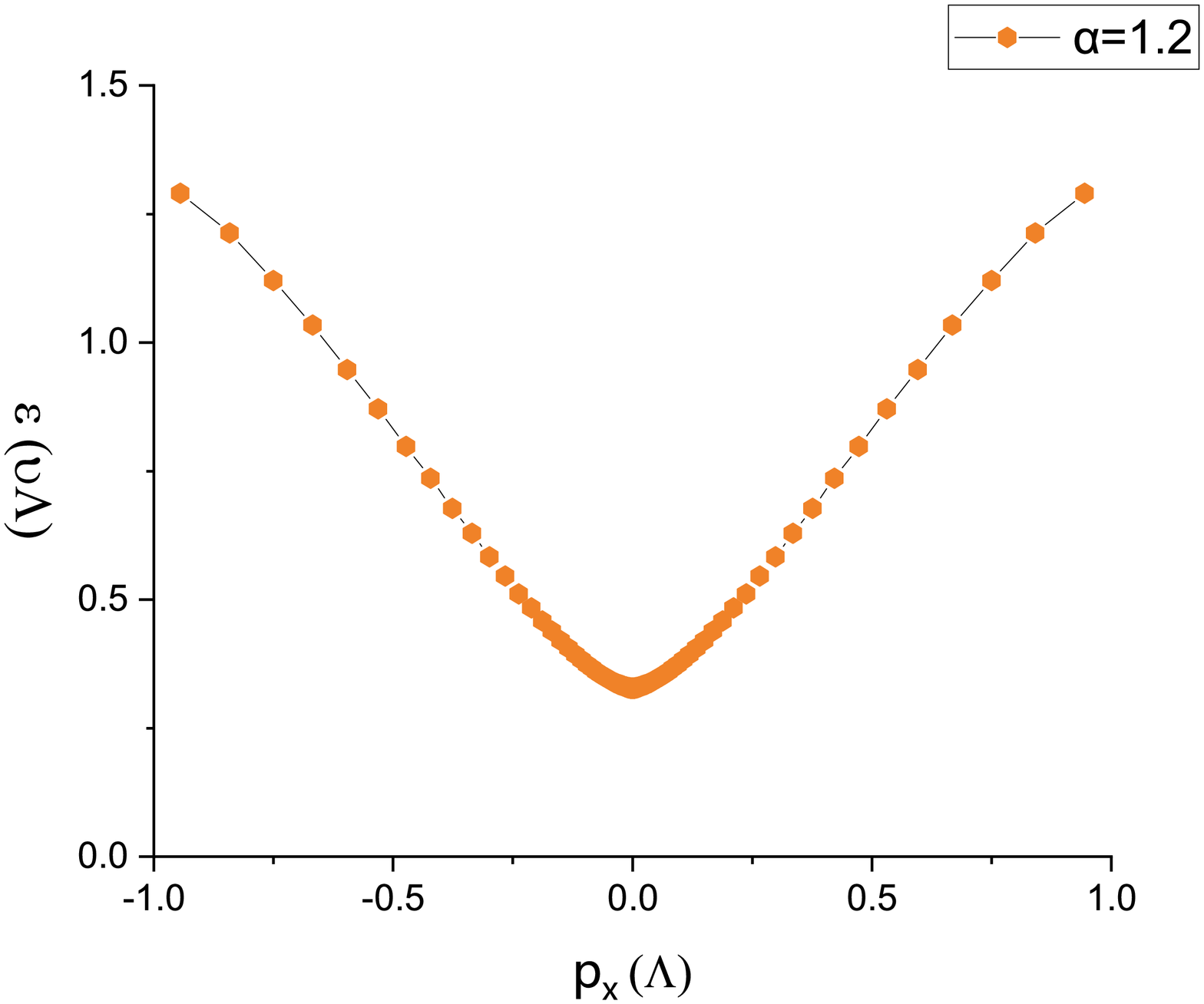}}
\subfigure[]{\label{Fig.sub.2}
\includegraphics[width=2.9in]{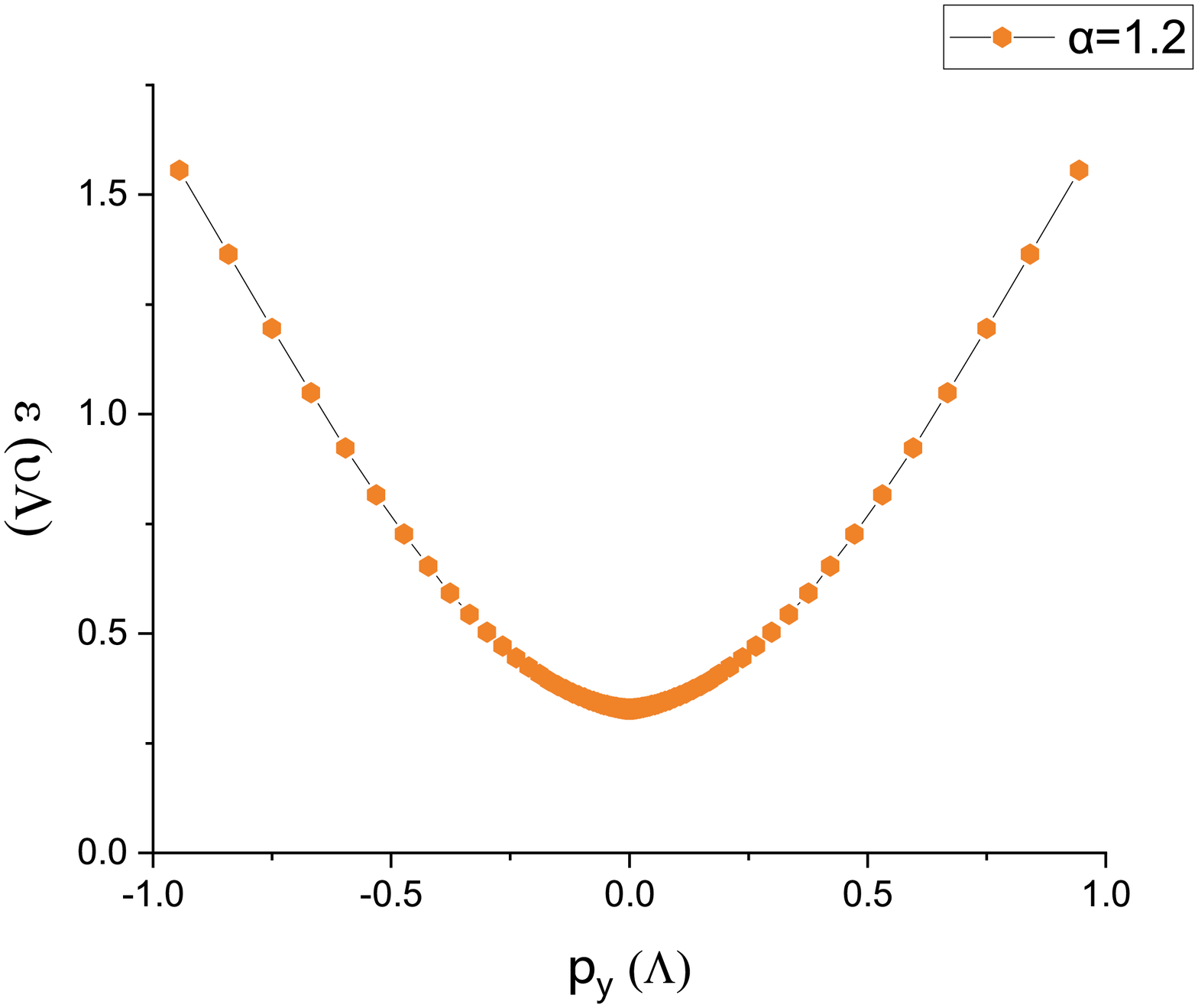}}
\caption{(a) The $p_{x}$-dependence of the renormalized fermion
dispersion $\varepsilon(p_{x},p_{y}) =
\pm\sqrt{A_1^2(p_{x},p_{y})p_x^2 + A_2^2(p_{x},p_{y})\beta^2p_y^4}$
obtained in the $p_{y}\rightarrow 0$ limit. (b) The
$p_{y}$-dependence of $\varepsilon(p_{x},p_{y})$ obtained in the
$p_{x}\rightarrow 0$ limit. Here, we set $\beta=1.0$ and
$\alpha=1.2$. In the case of strong coupling, the fermion
excitations exhibit an anisotropic linear dispersion due to the
strong renormalization caused by the Coulomb interaction.}
\label{Epxpy}
\end{figure}

\end{widetext}

The above results show that the function $A_{2}(p_x,p_y)\beta
p_{y}^{2}$ is linear in $p_{y}$ in the low-energy region. An
indication of such a behavior is that the semi-DSM state is actually
unstable against the Coulomb interaction since the originally
quadratic dispersion along the $y$-direction is strongly
renormalized and becomes a linear one. Consequently, the
renormalized dispersion of low-energy fermionic excitations of a 2D
semi-DSM resembles the cone-shaped dispersion of non-interacting 2D
Dirac fermions. By comparing Fig.~\ref{semi-Dirac}(a) to
Fig.~\ref{semi-Dirac}(b), one can see that the Coulomb interaction
renormalizes the dispersions of 2D Dirac fermions and 2D semi-Dirac
fermions very differently. Our non-perturbative results
qualitatively agree with that obtained by using the first-order
weak-coupling perturbative RG method \cite{Kotov21}, but appear to
be at odds with that obtained by the $1/N$-expansion approach
\cite{Isobe16, Cho16}.

We see from Fig.~\ref{A2pxpy}(c) that the renormalization function
$A_1(p_x,0)$ is nearly independent of $p_x$. Close to the limit of
$p_{x}\to 0$, $A_1(p_x,0)$ appears to be substantially enhanced. We
emphasize that such an enhancement is actually an artifact of IR
cutoff. There is always an IR cutoff for $p_{x}$ in carrying out
numerical computations. In our case, the IR cutoff is taken as
$10^{-5}\Lambda$. The contributions of the range of $p_{x} <
10^{-5}\Lambda$ to $A_{1}(p_x,0)$ are inevitably neglected. However,
as demonstrated in Ref.~\cite{Pan21}, the long-range nature of the
Coulomb interaction indicates that small-momentum contributions are
always larger than those of large momenta. If we reduce the IR
cutoff, the plateau always extends to lower momentum.

According to the results shown in Fig.~\ref{A2pxpy}(d),
$A_1(0,p_{y})$ is also nearly independent of $p_{y}$ for small
values of $\alpha$. But as $\alpha$ grows to fall into the
strong-coupling regime, $A_1(0,p_{y})$ becomes significantly
dependent on $p_{y}$. Based on the above results, we infer that the
fermion dispersion remains linear along the $x$-direction within a
broad range of $\alpha$ and momentum even when the impact of Coulomb
interaction is included.

In Fig.~\ref{Epxpy}(a) and Fig.~\ref{Epxpy}(b), we show how the
renormalized fermion spectrum, characterized by the function
$\varepsilon(p_{x},p_{y}) = \pm\sqrt{A_1^2(p_{x},p_{y})p_x^2 +
A_2^2(p_{x},p_{y})\beta^2p_y^4}$, depends on $p_{x}$ and $p_{y}$ for
a fixed value of $\alpha$, respectively. This spectrum manifests a
nearly linear dependence on momentum in both of the two directions
within a broad range of $p_{x}$ and $p_{y}$. There seems to be a
deviation from the standard linear behavior in the region of small
$p_{x,y}$. This deviation stems from the existence of a finite gap
at $p_{x,y}=0$. If such a gap is removed by tuning suitable external
parameters, the deviation from linear behavior would disappear.

\section{Summary and Discussion \label{Sec:Summary}}

In summary, we have performed a non-perturbative field theoretical
study of the renormalization of the dispersion of 2D semi-Dirac
fermions by incorporating the influence of long-range Coulomb
interaction. Making use of several exact identities, we derive the
self-closed DS integral equation of the full fermion propagator and
obtain the momentum-dependence of the renormalized fermion spectrum
based on the numerical solutions of such an equation. Our results
show that the originally quadratic dispersion of semi-Dirac fermions
is dramatically renormalized by the Coulomb interaction and changed
into a linear one, which is illustrated in Fig.~\ref{semi-Dirac}(a).
However, the originally linear dispersion remains linear after
renormalization.

Comparing to the RG approach, the DS equation approach has two major
advantages. Firstly, the full momentum-dependence of such physical
quantities as the renormalization functions $A_{1}(p_{x},p_{y})$ and
$A_{2}(p_{x},p_{y})$ can be obtained from the numerical solutions of
their integral equations. In contrast, the RG equations of various
model parameters depends merely on a varying scale $l = \ln
\left(\Lambda/E\right)$ where $E$ is either the energy or the
absolute quantity of the momentum. Secondly, the DS equation of
fermion propagator is exact and contains all the interaction-induced
effects. Hence the DS equation approach is valid for any values of
$\alpha$ and $N$.

In order to simplify numerical computations, we have neglected the
energy dependence of the renormalization functions
$A_{1}(p_{x},p_{y})$ and $A_{2}(p_{x},p_{y})$. Under such an
approximation, the quasiparticle resisue $Z$ cannot be calculated
since the renormalization function $A_{0}(p)=1$. Thus the results
presented in this paper cannot be used to judge whether the system
exhibit non-Fermi liquid behaviors. The computational time of
solving the complicated integral equations of
$A_{0}(p_{0},p_{x},p_{y})$, $A_{1}(p_{0},p_{x},p_{y})$, and
$A_{2}(p_{0},p_{x},p_{y})$ is much longer than that needed to solve
the integral equations of $A_{1}(p_{x},p_{y})$ and
$A_{2}(p_{x},p_{y})$. We wish to undertake such a task in a future
project.

\section*{ACKNOWLEDGEMENTS}

We thank Jing-Rong Wang and Zhao-Kun Yang for helpful discussions.
H.F.Z. is supported by the Natural Science Foundation of China
(Grants No. 12073026 and No. 11421303) and the Fundamental Research
Funds for the Central Universities.


\begin{thebibliography}{99}

\bibitem{CastroNeto}
A. H. Castro Neto, F. Guinea, N. M. R. Peres, K. S. Novoselov, and
A. K. Geim, Rev. Mod. Phys. {\bf 81}, 109 (2009).

\bibitem{Kotov12}
V. N. Kotov, B. Uchoa, V. M. Pereira, F. Guinea, and A. H. Castro
Neto, Rev. Mod. Phys. {\bf 84}, 1067 (2012).

\bibitem{Vafek14}
O. Vafek and A. Vishwanath, Annu. Rev. Condens. Matter Phys. {\bf
5}, 83 (2014).

\bibitem{Hirata21}
M. Hirata, A. Kobayashi, C. Berthier, and K. Kanoda, Rep. Prog.
Phys. {\bf 84}, 036502 (2021).

\bibitem{Shankar}
R. Shankar, Rev. Mod. Phys. {\bf66}, 129 (1994).

\bibitem{Coleman}
P. Coleman, \emph{Introduction to Many-Body Physics} (Cambridge
University Press, Cambridge, 2015).

\bibitem{Gonzalez94}
J. Gonz\'{a}lez, F. Guinea, and M. A. H. Vozmediano, Nucl. Phys. B
{\bf 424}, 595 (1994).

\bibitem{Elias11}
D. C. Elias, R. V. Gorbachev, A. S. Mayorov, S. V. Morozov, A. A.
Zhukov, P. Blake, L. A. Ponomarenko, I. V. Grigorieva, K. S.
Novoselov, F. Guinea, and A. K. Geim, Nat. Phys. {\bf 7}, 701
(2011).

\bibitem{Lanzara11}
D. A. Siegel, C.-H. Park, C. Hwang, J. Deslippe, A. V. Fedorov, S.
G. Louie, and A. Lanzara, Proc. Natl. Acad. Sci. U.S.A. {\bf 108},
11365 (2011).

\bibitem{Chae12}
J. Chae, S. Jung, A. F. Young, C. R. Dean, L. Wang, Y. Gao, K.
Watanabe, T. Taniguchi, J. Hone, K. L. Shepard, P. Kim, N. B.
Zhitenev, and J. A. Stroscio , Phys. Rev. Lett. {\bf 109}, 116802
(2012).

\bibitem{Pan21}
X.-Y. Pan, Z.-K. Yang, X. Li, and G.-Z. Liu, Phys. Rev. B {\bf 104},
085141 (2021).

%\bibitem{Hasegawa06}
%Y. Hasegawa, R. Konno, H. Nakano, and M. Kohmoto, Phys. Rev. B {\bf
%74}, 033413 (2006).

\bibitem{Montambaux09A}
G. Montambaux, F. Pi\'{e}chon, J.-N. Fuchs, and M. O. Goerbig, Eur.
Phys. J. B {\bf 72}, 509 (2009).

\bibitem{Montambaux09B}
G. Montambaux, F. Pi\'{e}chon, J.-N. Fuchs, and M. O. Goerbig, Phys.
Rev. B {\bf 80}, 153412 (2009).

\bibitem{Lim12}
L.-K. Lim, J.-N. Fuchs, and G. Montambaux, Phys. Rev. Lett. {\bf
108}, 175303 (2012).

\bibitem{Bellec13}
M. Bellec, U. Kuhl, G. Montambaux, and F. Mortessagne, Phys. Rev.
Lett. {\bf 110}, 033902 (2013).

\bibitem{Dietl08}
P. Dietl, F. Pi\'{e}chon, and G. Montambaux, Phys. Rev. Lett. {\bf
100}, 236405 (2008).

\bibitem{Goerbig08}
M. O. Goerbig, J.-N. Fuchs, G. Montambaux, and F. Pi\'{e}chon, Phys.
Rev. B {\bf 78}, 045415 (2008).

\bibitem{Kobayashi07}
A. Kobayashi, S. Katayama, Y. Suzumura, and H. Fukuyama, J. Phys.
Soc. Jpn. {\bf 76}, 034711 (2007).

\bibitem{Kobayashi11}
A. Kobayashi, Y. Suzumura, F. Pi\'{e}chon, and G. Montambaux, Phys.
Rev. B {\bf 84}, 075450 (2011).

\bibitem{Pardo09}
V. Pardo and W. E. Pickett, Phys. Rev. Lett. {\bf 102}, 166803
(2009).

\bibitem{Pardo10}
V. Pardo and W. E. Pickett, Phys. Rev. B {\bf 81}, 035111 (2010).

\bibitem{Banerjee09}
S. Banerjee, R. R. P. Singh, V. Pardo, and W. E. Pickett, Phys. Rev.
Lett. {\bf 103}, 016402 (2009).

\bibitem{Kim15}
J. Kim, S. S. Baik, S. H. Ryu, Y. Sohn, S. Park, B.-G. Park, J.
Denlinger, Y. Yi, H. J. Choi, and K. S. Kim, Science {\bf 349}, 723
(2015).

\bibitem{Wunsch08}
B. Wunsch, F. Guinea, and F. Sols, New. J. Phys. {\bf 10}, 103027
(2008).

\bibitem{Tarruell12}
L. Tarruell, D. Greif, T. Uehlinger, G. Jotzu, and T. Esslinger,
Nature {\bf 483}, 302 (2012).

\bibitem{Isobe16}
H. Isobe, B.-J. Yang, A. Chubukov, J. Schmalian, and N. Nagaosa,
Phys. Rev. Lett. {\bf 116}, 076803 (2016).

\bibitem{Cho16}
G. Y. Cho and E.-G. Moon, Sci. Rep. {\bf 6}, 19198 (2016).

\bibitem{Kotov21}
V. N. Kotov, B. Uchoa, and O. P. Sushkov, Phys. Rev. B {\bf 103},
045403 (2021).

\bibitem{Schmalian18}
J. M. Link, B. N. Narozhny, E. I. Kiselev, and J. Schmalian, Phys.
Rev. Lett. {\bf 120}, 196801 (2018).

\bibitem{Wang17}
J.-R. Wang, G.-Z. Liu, and C.-J. Zhang, Phys. Rev. B {\bf 95},
075129 (2017).

\bibitem{Uchoa17}
B. Uchoa and K. Seo, Phys. Rev. B {\bf 96}, 220503(R) (2017).

\bibitem{Uchoa19}
M. D. Uryszek, E. Christou, A. Jaefari, F. Kr\"{u}ger, and B. Uchoa,
Phys. Rev. B {\bf 100}, 155101 (2019).

\bibitem{Roy18}
B. Roy and M. S. Foster, Phys. Rev. X {\bf 8}, 011049 (2018).

\bibitem{Liu21}
G.-Z. Liu, Z.-K. Yang, X.-Y. Pan, and J.-R. Wang, Phys. Rev. B {\bf
103}, 094501 (2021).

%\bibitem{Itzykson}
%C. Itzykson and J.-B. Zuber, \emph{Quantum Field Theory}
%(McGraw-Hill, New York, 1980).

%\bibitem{Appelquist86}
%T. W. Appelquist, M. Bowick, D. Karabali, and L. C. R. Wijewardhana,
%Phys. Rev. D {\bf 33}, 3704 (1986).

\end{thebibliography}
\end{document}